\documentclass[a4paper,10pt]{article}
\usepackage[margin=1.3cm]{geometry}
\usepackage{comment}
\usepackage{amssymb,extarrows,graphicx,subfigure,setspace}
\usepackage{cite}
\usepackage{slashed}
\usepackage{tensor}
\usepackage[toc,page]{appendix}
\usepackage{color}
\usepackage{physics}
\usepackage{hyperref}
\usepackage{dirtytalk}
\hypersetup{colorlinks=true, linkcolor=blue, citecolor=red, linktoc=page}
\makeatother

\usepackage{float}
\usepackage{textcomp}
\usepackage{amsmath}
\newmuskip\pFqmuskip

\newcommand*\pFq[6][8]{%
  \begingroup 
  \pFqmuskip=#1mu\relax
  \mathcode`\,=\string"8000
  \begingroup\lccode`\~=`\,
  \lowercase{\endgroup\let~}\pFqcomma
  {}_{#2}F_{#3}{\left[\genfrac..{0pt}{}{#4}{#5};#6\right]}%
  \endgroup
}
\newcommand{\pFqcomma}{\mskip\pFqmuskip}

\usepackage{mathrsfs}
\usepackage{soul}
\usepackage{hyperref}

\newcommand{\be}{\begin{equation}}
\newcommand{\bea}{\begin{eqnarray}}
\newcommand{\eea}{\end{eqnarray}}
\newcommand{\ba}{\begin{array}}
\newcommand{\ea}{\end{array}}
\newcommand{\ee}{\end{equation}}
\newcommand{\bes}{\begin{equation*}}
\newcommand{\beas}{\begin{eqnarray*}}
\newcommand{\eeas}{\end{eqnarray*}}
\newcommand{\bas}{\begin{array*}}
\newcommand{\eas}{\end{array*}}
\newcommand{\ees}{\end{equation*}}

\numberwithin{equation}{section}

\begin{document}
\color{black}
\begin{center}
\Large{\bf Non-extensive Entropy and Holographic Thermodynamics: Topological Insights}\\
\small \vspace{0.7cm}

\small {\bf Saeed Noori Gashti $^{\dag}$\footnote {Email:~~~saeed.noorigashti@stu.umz.ac.ir; saeed.noorigashti70@gmail.com}},\quad
{\bf B. Pourhassan $^{\dag}$\footnote {Email:~~~b.pourhassan@du.ac.ir, b.pourhassan@candqrc.ca}},\quad\\
 \vspace{0.5cm}$^{\dag}${School of Physics, Damghan University, P. O. Box 3671641167, Damghan, Iran}\\
\small \vspace{1cm}
\end{center}
\begin{abstract}
In this paper, we delve into the thermodynamic topology of AdS Einstein-Gauss-Bonnet black holes, employing non-extensive entropy formulations such as Barrow, Rényi, and Sharma-Mittal entropy within two distinct frameworks: bulk boundary and restricted phase space (RPS) thermodynamics. Our findings reveal that in the bulk boundary framework, the topological charges, influenced by the free parameters and the Barrow non-extensive parameter $(\delta)$, exhibit significant variability. Specifically, we identify three topological charges $(\omega = +1, -1, +1)$. When the parameter $\delta$ increases to 0.9, the classification changes, resulting in two topological charges $(\omega = +1, -1)$. When $\delta$ is set to zero, the equations reduce to the Bekenstein-Hawking entropy structure, yielding consistent results with three topological charges. Additionally, setting the non-extensive parameter $\lambda$ in Rényi entropy to zero increases the number of topological charges, but the total topological charge remains (W = +1). The presence of the Rényi non-extensive parameter alters the topological behavior compared to the Bekenstein-Hawking entropy. Sharma-Mittal entropy shows different classifications and the various numbers of topological charges influenced by the non-extensive parameters $\alpha$ and $\beta$. When $\alpha$ and $\beta$ have values close to each other, three topological charges with a total topological charge $(W = +1)$ are observed. Varying one parameter while keeping the other constant significantly changes the topological classification and number of topological charges. In contrast, the RPS framework demonstrates remarkable consistency in topological behavior. Under all conditions and for all free parameters, the topological charge remains $(\omega = +1)$ with the total topological charge $(W = +1)$. This uniformity persists even when reduced to Bekenstein-Hawking entropy, suggesting that the RPS framework provides a stable environment for studying black hole thermodynamics across different entropy models. These findings underscore the importance of considering various entropy formulations and frameworks to gain a comprehensive understanding of black hole thermodynamics.\\\\
Keywords; Thermodynamic topology, Nonextensive entropy, bulk boundary, Restricted phase space\\
\end{abstract}
\tableofcontents
\newpage
\section{Introduction}
The area theorem of black holes\cite{900}, proposed by Stephen Hawking, states that the total horizon area of black holes cannot decrease over time during any physical process that adheres to the laws of classical physics. This theorem suggests that black holes possess thermodynamic properties, as they mirror the behavior of entropy in thermodynamic systems. Jacob Bekenstein further developed this idea by proposing that the entropy of a black hole is proportional to the area of its event horizon. This relationship, known as the Bekenstein-Hawking entropy, highlights a profound connection between the geometry of black holes and thermodynamic entropy\cite{901,902}. The analogy between black hole thermodynamics and classical thermodynamics was further solidified by Stephen Hawking's discovery of Hawking radiation. This phenomenon occurs due to quantum effects near the event horizon, causing black holes to emit thermal radiation. As a result, black holes can be assigned a temperature, known as the Hawking temperature, which is inversely proportional to their mass\cite{903,904,905}.\\

Recently introduced a groundbreaking method to examine the topological charge of black holes. This method interprets black hole solutions as topological defects within the thermodynamic parameter space. By employing the generalized off-shell free energy, they categorized black holes based on their topological charge, which is determined by the winding numbers of these defects. Black holes with positive winding numbers are considered locally stable, while those with negative winding numbers are deemed locally unstable. This innovative approach offers a new perspective on the thermodynamic stability of black holes and provides valuable insights into phase transitions and critical phenomena in black hole thermodynamics. It has been applied to various black holes, including those in anti-de Sitter (AdS) spacetime, uncovering new types of critical points and phase behaviors\cite{a19,a20}.

The topological method for black hole thermodynamics has become popular due to its straightforwardness in examining thermodynamic properties. It has been utilized to investigate the Hawking-Page phase transition of Schwarzschild-AdS black holes and their holographic counterparts, which relate to the confinement-deconfinement transition in gauge theories. Quantum gravity corrections, expressed through higher-derivative terms, have been studied for black holes in Einstein-Gauss-Bonnet and Lovelock gravity. These corrections shed light on the behavior of black holes in higher-dimensional spacetimes and the effects of quantum gravity. Although these studies mainly focus on static black holes, the topological approach has also been extended to rotating black holes, offering significant insights into their thermodynamic properties, stability, topological classification, and topological photon spheres\cite{20a,21a,22a,23,24,25,26,27,28,29,31,33,34,35,37,38,38a,38b,38c,39,40,41,42,43,44,44c,44d}.\\

In this article, we aim to explore the topology of holographic thermodynamics using non-extensive entropies such as Barrow, Rényi, and Sharma-Mittal entropy. Our objective is to identify the topological class of these black holes and compare it with the Bekenstein-Hawking entropy. Non-extensive entropy, often linked with Tsallis entropy, is a generalization of the traditional Boltzmann-Gibbs entropy. This concept was introduced by Tsallis to address systems where the conventional assumptions of extensive entropy do not apply. In classical thermodynamics, entropy is extensive, meaning it scales linearly with the system's size. However, many physical systems exhibit non-extensive behavior due to long-range interactions, fractal structures, or other complexities\cite{aa,bb,cc,dd,ee,ff}. Non-extensive entropy has been applied to various astrophysical phenomena, including the distribution of stellar objects and the dynamics of galaxy clusters. It aids in modeling systems where gravitational interactions are long-range and cannot be described by extensive entropy. Non-extensive entropy extends information theory concepts to systems with non-standard probability distributions. It is utilized in coding theory, data compression, and the analysis of complex networks\cite{aa,bb,cc,dd,ee,ff}.\\

Holographic thermodynamics is a framework that applies the principles of holography to the study of black hole thermodynamics. This approach often involves the AdS/CFT correspondence, which posits a relationship between a gravitational theory in an anti-de Sitter (AdS) space and a conformal field theory (CFT) on its boundary. This duality allows physicists to study complex gravitational systems using quantum field theories' simpler, well-understood properties. Thus, one can study two spaces with features such as bulk-boundary correspondence and restricted phase space. Bulk-boundary correspondence is a principle that connects the properties of a bulk system (like a black hole in AdS space) with those of its boundary (the CFT). This correspondence is crucial in understanding topological phases of matter and has applications in condensed matter physics and high-energy physics. It essentially states that the behavior of a system's boundary can reveal information about the bulk properties\cite{701,702,703,704,705,706,707,708,709,710,711,712,713,714,715}. Restricted phase space thermodynamics is a newer formalism that modifies traditional black hole thermodynamics by fixing certain parameters, such as the AdS radius, as constants. This approach eliminates the need for pressure and volume as thermodynamic variables, instead using the central charge and chemical potential. This formalism maintains the Euler relation equation, providing a consistent framework for studying black hole thermodynamics\cite{701,702,703,704,705,706,707,708,709,710,711,712,713,714,715}.\\
Based on these explanations, we will organize the article as follows:

In Section 2, we will delve into the concept of Nonextensive Entropy. This section will cover some models and formulas and their applications in various physical systems. We will overview the Nonextensive Entropy, associated with Barrow, Rényi, and Sharma-Mittal, which extends the traditional Boltzmann-Gibbs framework to accommodate systems with long-range interactions, fractal structures, and other complexities that exhibit non-extensive behavior. Section 3 will explain the thermodynamic topology using the generalized Helmholtz free energy method. We will discuss how this method allows us to classify black holes based on their topological charge, determined by the winding numbers of topological defects in the thermodynamic parameter space. This section will also overview the implications of this classification for understanding the stability and phase transitions of black holes.
In Section 4, we will provide a comprehensive overview of the black hole model within the frameworks of bulk-boundary correspondence and restricted phase space. This section will include detailed calculations and discussions on the thermodynamic topology of the model, with a particular focus on non-extensive entropies such as Barrow, Rényi, and Sharma-Mittal entropy. We will examine how these entropies influence the thermodynamic properties and stability of black holes, and compare them with the traditional Bekenstein-Hawking entropy.
Finally, Section 5 will present our conclusions and summarize the key findings of our study. We will reflect on the insights gained from our exploration of non-extensive entropies and thermodynamic topology, and discuss the broader implications of our results for the field of black hole thermodynamics. This section will also suggest potential directions for future research, building on the foundations laid by our work.
\section{Non-extensive Entropy}
Non-extensive entropy is an extension of the traditional Boltzmann-Gibbs entropy, introduced by Constantino Tsallis. This concept is particularly useful for systems that exhibit non-linearity and a strong dependence on initial conditions. Unlike Boltzmann-Gibbs entropy, which assumes that entropy scales linearly with the size of the system, non-extensive entropy can handle systems where this linearity does not hold. This makes it applicable to a wide range of fields, including theoretical physics, cosmology, and statistical mechanics. It is especially relevant for systems with long-range interactions, fractal structures, or memory effects\cite{a33}.
\subsection{Rényi entropy}
Rényi entropy is one form of non-extensive entropy that has been used to study black hole thermodynamics. It is defined by a parameter that adjusts the degree of non-extensiveness. This parameter must fall within a specific range to ensure the entropy function remains well-defined. When applied to black holes, Rényi entropy provides a framework for understanding their thermodynamic properties in a way that generalizes the traditional Boltzmann-Gibbs statistics\cite{a25,a26,a27}.
\begin{equation}\label{N1}
\begin{split}
S_R = \frac{1}{\lambda} \ln(1 + \lambda S_{BH})
\end{split}
\end{equation}
The parameter $(\lambda)$ in non-extensive entropy plays a crucial role in defining the entropy function. For the entropy function to remain well-defined, $(\lambda)$ must lie within the range $(-\infty < \lambda < 1)$. Values outside this range make the entropy function convex and thus ill-defined. In the context of black hole thermodynamics using Rényi statistics, the entropy $(S_R)$ is properly defined when $(\lambda)$ is between 0 and 1. Within this interval, $(\lambda)$ exhibits favorable thermodynamic properties, as demonstrated in recent studies. Notably, as the Rényi parameter $(\lambda)$ approaches zero, the generalized off-shell free energy converges to the classical Boltzmann-Gibbs statistics.
\subsection{Sharma-Mittal entropy}
Another important form of non-extensive entropy is the Sharma-Mittal entropy, which generalizes both Rényi and Tsallis entropies. This entropy has been particularly insightful in cosmological studies, such as describing the accelerated expansion of the universe by effectively utilizing vacuum energy. Although non-extensive entropies have been used to study black holes, the Sharma-Mittal entropy has not yet been extensively applied in this context. This presents an opportunity to explore the thermodynamic properties of black holes using Sharma-Mittal entropy, considering them as strongly coupled gravitational systems\cite{a28,a29,a30},
\begin{equation}\label{N2}
\begin{split}
S_{SM} = \frac{1}{\alpha} \left( (1 + \beta S_T)^\frac{\alpha}{\beta} - 1 \right).
\end{split}
\end{equation}
In this context, $S_T$ denotes the Tsallis entropy, which is derived from the horizon area $(A = 4\pi r^2)$, where $r$ is the radius of the black hole's event horizon. The parameters $\alpha$ and $(\beta)$ are adjustable and need to be calibrated using observational data. Interestingly, when $\alpha$ approaches zero, the Sharma-Mittal entropy simplifies to the Rényi entropy. Similarly, when $\alpha$ equals $(\beta)$, it reduces to the Tsallis entropy.
\subsection{Barrow entropy}
Barrow entropy is another intriguing concept that arises from quantum gravity effects. These effects can deform the surface of a black hole, resulting in a fractal structure. This deformation modifies the black hole's entropy, leading to what is known as Barrow entropy. The extent of these deformations is measured by a parameter, and depending on its value, the entropy can range from the traditional Bekenstein-Hawking entropy (with no fractal structure) to a highly deformed, complex fractal structure\cite{a31,a32},
\begin{equation}\label{N3}
\begin{split}
S_B = \left( \frac{A}{A_{PI}} \right)^{\frac{1 + \delta}{2}}
\end{split}
\end{equation}
In this context, $A$ denotes the area of the black hole's event horizon, while $A_{PI}$ refers to the Planck area. The parameter $\delta$ quantifies the degree of quantum gravity-induced deformations on the event horizon, ranging from 0 to 1. When $\delta$ is zero, the entropy reverts to the Bekenstein-Hawking form, indicating no fractal deformation. This scenario aligns with the conventional analysis of Reissner-Nordström AdS black holes without any fractal modifications. On the other hand, a $\delta$ value of one represents the maximum deformation, resulting in a highly complex fractal structure of the event horizon.
In summary, non-extensive entropies like Rényi, Sharma-Mittal, and Barrow entropies provide powerful tools for exploring the thermodynamic properties of black holes. They offer new perspectives and insights, particularly in systems where traditional thermodynamic assumptions do not apply. These concepts continue to expand our understanding of black holes and their role in the universe.
\section{Thermodynamic topology}
Recent advancements have introduced innovative methods for analyzing and computing critical points and phase transitions in black hole thermodynamics. One prominent approach is the topological method, which leverages Duan’s topological current $\phi$-mapping theory to adopt a topological perspective in thermodynamics\cite{a19,a20}.
To investigate the thermodynamic properties of black holes, various quantities such as mass and temperature are used to describe the generalized free energy. Given the relationship between mass and energy in black holes, the generalized free energy function is expressed as a standard thermodynamic function. The Euclidean time period $\tau$ and its inverse, the temperature $T$, are key components in this formulation. The generalized free energy is considered on-shell only when $\tau$ equals the inverse of the Hawking temperature\cite{a19,a20}. A vector $\phi$ is constructed to facilitate this analysis, with components derived from the partial derivatives of the generalized free energy. The direction of this vector is significant, as it points outward at specific angular positions, indicating the ranges for the horizon radius and angular coordinates. Using Duan's $\phi$-mapping topological current theory, a topological current can be defined, which is conserved according to Noether's theorem.
To determine the topological number, the topological current is reformulated, incorporating the Jacobi tensor. This tensor simplifies to the standard Jacobi form under certain conditions, and the conservation equation reveals that the topological current is non-zero only at specific points. Through detailed calculations, the topological number or total charge $W$ can be expressed, involving the Hopf index and the sign of the topological current at zero points. The winding number, which is independent of the region's shape, directly relates to black hole stability. A positive winding number corresponds to a stable black hole state, while a negative winding number indicates instability.
This topological approach provides a robust framework for understanding the stability and phase transitions of black holes, offering new insights into their thermodynamic behavior. So The generalized free energy is determined as\cite{a19,a20},
\begin{equation}\label{F1}
\mathcal{F} = M - \frac{S}{\tau},
\end{equation}
In this context, $\tau$ signifies the Euclidean time period, and its inverse, $T$, represents the temperature of the system. The generalized free energy is considered on-shell only when $\tau$ matches the inverse of the Hawking temperature. To facilitate this analysis, a vector $(\phi)$ is constructed with components derived from the partial derivatives as follows,
\begin{equation}\label{F2}
\phi = \left(\frac{\partial \mathcal{F}}{\partial r_{H}}, -\cot \Theta \csc \Theta \right).
\end{equation}
In this scenario, $(\phi^{\Theta}$ becomes infinite, and the vector points outward at the angles $(\Theta = 0)$ and $(\Theta = \pi)$. The permissible ranges for the horizon radius $(r_{H})$ and the angle $(\Theta)$ are from 0 to infinity and from 0 to $(\pi)$, respectively. By applying Duan's $(\phi)$-mapping topological current theory, we can define a topological current as follows:
\begin{equation}\label{F3}
j^{\mu} = \frac{1}{2\pi} \varepsilon^{\mu\nu\rho} \varepsilon_{ab} \partial_{\nu} n^{a} \partial_{\rho} n^{b}, \quad \mu, \nu, \rho = 0, 1, 2,
\end{equation}
In this formulation, $n$ is defined as $(n^1, n^2)$, where $(n^1 = \frac{\phi^r}{|\phi|})$ and $(n^2 = \frac{\phi^\Theta}{|\phi|})$. According to the conservation equation, the current $(j^{\mu})$ is non-zero exclusively at the points where $(\phi = 0)$. After performing the necessary calculations, the topological number or total charge $W$ can be determined as follows:
\begin{equation}\label{F4}
W = \int_{\Sigma} j^{0} d^2 x = \sum_{i=1}^{n} \beta_{i} \eta_{i} = \sum_{i=1}^{n} \omega_{i}.
\end{equation}
In this context, $(\beta_i)$ represents the positive Hopf index, which counts the number of loops made by the vector $(\phi^a)$ in the $(\phi)$-space when $(x^\mu)$ is close to the zero point $(z_i)$. Meanwhile, $(\eta_i)$ is defined as the sign of $(j^0(\phi/x)_{z_i})$, which can be either +1 or -1. The term $(\omega_i)$ denotes the winding number associated with the (i)-th zero point of $(\phi)$ within the region $(\Sigma)$
\section{AdS Einstein-Gauss-Bonnet black holes}
AdS Einstein-Gauss-Bonnet (EGB) black holes are solutions in anti-de Sitter space that incorporate a higher-order curvature correction term. These black holes emerge from the EGB equations with a negative cosmological constant. Depending on the mass, charge, and Gauss-Bonnet coupling constant, these black holes can have either one or two horizons. They exhibit a Hawking temperature, entropy, and electrical potential, all of which comply with the first law of thermodynamics. A distinctive feature of AdS EGB black holes is their phase transition behavior. They can transition from a small black hole to a large black hole, or vice versa, when the temperature or pressure reaches a critical value. This phase transition is different from the liquid-gas phase transition seen in Van der Waals fluids and is influenced by the sign and magnitude of the Gauss-Bonnet coupling constant. The thermodynamics of AdS EGB black holes focuses on understanding the properties and behavior of these solutions in relation to their mass, temperature, entropy, heat capacity, and free energy. The EGB theory in (D) dimensions is described by the action\cite{950,951},
\begin{equation}\label{M1}
S = \frac{1}{16\pi} \int d^D x \sqrt{-g} (R + a L),
\end{equation}
where $L$ is defined as,
\begin{equation}\label{M2}
L = R^2 - 4R_{\mu\nu}R^{\mu\nu} + R_{\mu\nu\rho\sigma}R^{\mu\nu\rho\sigma}.
\end{equation}
In this notation, $R$ is the Ricci scalar, $R_{\mu\nu}$ is the Ricci tensor, and $R_{\mu\nu\rho\sigma}$ is the Riemann tensor. The metric $g_{\mu\nu}$ has a determinant denoted by $g$. The Gauss-Bonnet term does not affect the dynamics in four dimensions $(D = 4)$ because its integral is a topological invariant. However, the coupling constant can be adjusted by scaling it as,
\begin{equation}\label{M3}
a \rightarrow \frac{a}{D - 4}
\end{equation}
The metric of an AdS EGB black hole with spherical symmetry can be expressed as,
\begin{equation}\label{M4}
ds^2 = -f(r) dt^2 + \frac{dr^2}{f(r)} + r^2 d\Omega^2_{D-2},
\end{equation}
where $d\Omega^2_{D-2}$ is the metric on a unit $(D-2)$-sphere. The metric function $f(r)$ for the 4D AdS EGB black hole is given by,
\begin{equation}\label{M4}
f(r) = 1 + \frac{r^2}{2a} \left(1 - \sqrt{1 + 4a \left(\frac{2M}{r^3} - \frac{q^2}{r^4} - \frac{1}{l^2}\right)}\right).
\end{equation}
Here, $M$ is the mass parameter, $\Lambda$ is the cosmological constant, and $a$ is the Gauss-Bonnet coupling constant. The temperature of the black hole is determined by,
\begin{equation}\label{M5}
T = \frac{f'(r_h)}{4\pi}.
\end{equation}
This comprehensive framework allows for the exploration of the thermodynamic properties of AdS EGB black holes, providing insights into their stability, phase transitions, and overall behavior in higher-dimensional spacetimes.
\subsection{Bulk boundary thermodynamics}
In this subsection, we consider the 4D AdS Einstein-Gauss-Bonnet (EGB) black hole. The metric function for this black hole is given by,
\begin{equation}\label{MB1}
f(r) = 1 + \frac{r^2}{2a} \left(1 - \sqrt{1 + 4a \left(\frac{2MG}{r^3} - \frac{q^2G}{r^4} - \frac{1}{l^2}\right)}\right).
\end{equation}
The radius of the anti-de Sitter (AdS) space and the entropy for this black hole are expressed as,
\begin{equation}\label{MB2}
l = \frac{1}{4}\sqrt{\frac{6}{P G \pi}},\quad S = \frac{r_h^2 \pi}{G} + 4 \ln \left(\frac{r_h}{\sqrt{a}}\right)a \pi.
\end{equation}
The Hawking temperature of the AdS EGB black hole, rewritten according to the relevant equation, is,
\begin{equation}\label{MB3}
T = \frac{8P G \pi r_h^4 - q^2 G + r_h^2 - a}{4 r_h^3 \pi + 8 r_h a \pi},
\end{equation}
The variable cosmological constant for this model is given by,
\begin{equation}\label{MB4}
G = -\frac{-r_h^4 + 5 a r_h^2 + 2 a^2}{8P \pi r_h^6 + 48P \pi a r_h^4 + 3q^2 r_h^2 + 2 a q^2}.
\end{equation}
\subsubsection{Thermodynamic topology within Barrow statistics}
\begin{figure}[h!]
 \begin{center}
 \subfigure[]{
 \includegraphics[height=4cm,width=4cm]{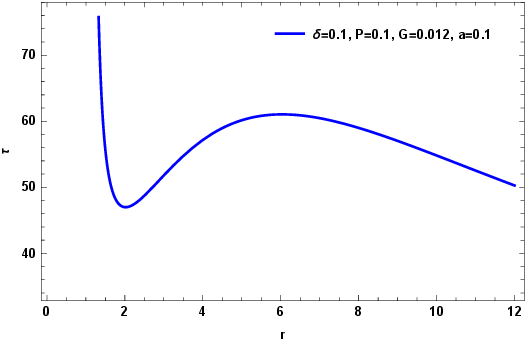}
 \label{1a}}
 \subfigure[]{
 \includegraphics[height=4cm,width=4cm]{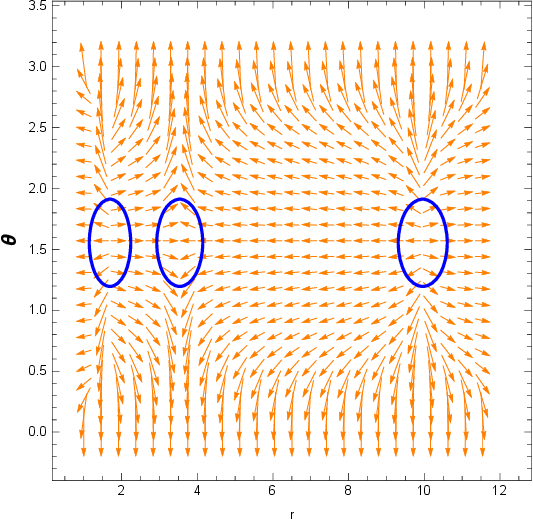}
 \label{1b}}
 \subfigure[]{
 \includegraphics[height=4cm,width=4cm]{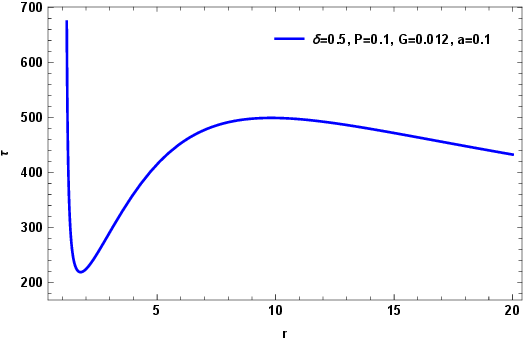}
 \label{1c}}
 \subfigure[]{
 \includegraphics[height=4cm,width=4cm]{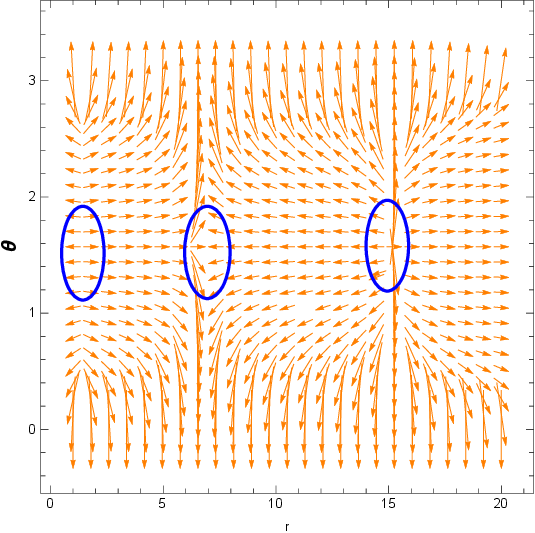}
 \label{1d}}\\
 \subfigure[]{
 \includegraphics[height=4cm,width=4cm]{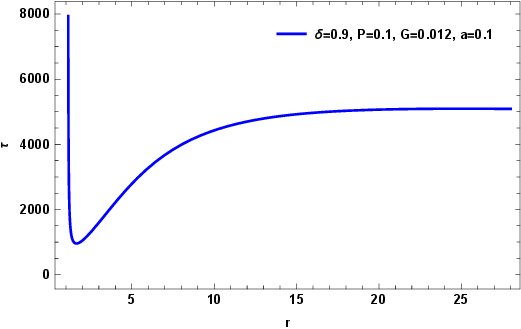}
 \label{1e}}
 \subfigure[]{
 \includegraphics[height=4cm,width=4cm]{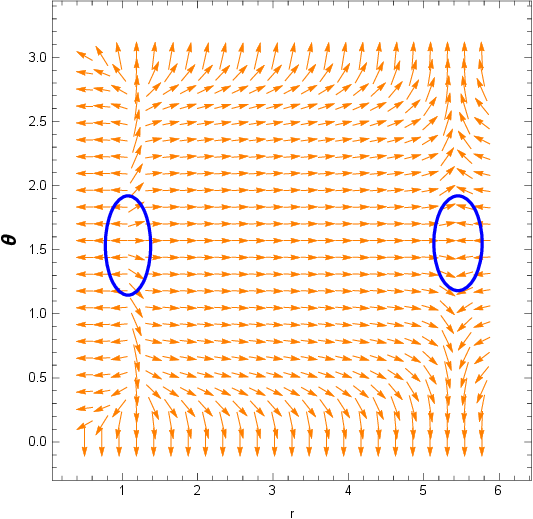}
 \label{1f}}
 \subfigure[]{
 \includegraphics[height=4cm,width=4cm]{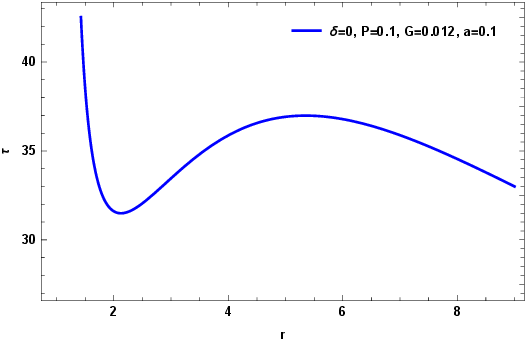}
 \label{1g}}
 \subfigure[]{
 \includegraphics[height=4cm,width=4cm]{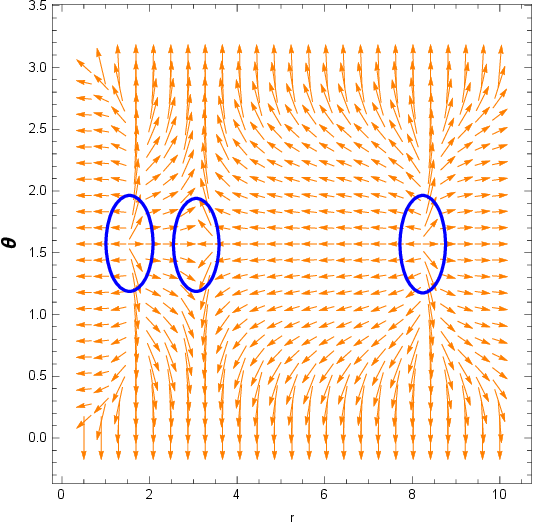}
 \label{1h}}
  \caption{\small{The curve described by Eq. (\ref{MBB3}) is illustrated in Figs. (\ref{1a}), (\ref{1c}), (\ref{1e}), and (\ref{1g}). In Figs. (\ref{1b}), (\ref{1d}), (\ref{1f}), and (\ref{1h}), the zero points (ZPs) are located at coordinates $(r, \theta)$ on the circular loops, corresponding to the nonextensive parameter $\delta$.}}
 \label{m1}
 \end{center}
 \end{figure}

Here, we explore the thermodynamic topology within the framework of Barrow entropy for bulk boundary thermodynamics. Using Eqs. (\ref{N1}), (\ref{F1}), and (\ref{MB1}), we derive the function $\mathcal{F}$. Consequently, we can calculate $\phi^{r_h}$ and $\phi^{\theta }$ with respect to Eqs. (\ref{F2}) as follows,
\begin{equation}\label{MBB1}
\phi^r=-\frac{\tau  \left(a+G \left(Q^2-8 \pi  P r^4\right)-r^2\right)+2 \pi  (\delta +2) r \left(2 a G+r^2\right) \left(4 \pi  a \ln \left(\frac{r}{\sqrt{a}}\right)+\frac{\pi  r^2}{G}\right)^{\delta /2}}{2 G r^2 \tau }
\end{equation}
and
\begin{equation}\label{MBB2}
\phi^{\theta }=-\frac{\cot (\theta )}{\sin (\theta )}
\end{equation}
Additionally, we determine $\tau$ as follows,
\begin{equation}\label{MBB3}
\tau =\frac{2 \pi  (\delta +2) r \left(2 a G+r^2\right) \left(4 \pi  a \ln \left(\frac{r}{\sqrt{a}}\right)+\frac{\pi  r^2}{G}\right)^{\delta /2}}{-a+8 \pi  G P r^4-G Q^2+r^2}
\end{equation}
In our study, we investigate the thermodynamic topology of AdS Einstein-Gauss-Bonnet black holes using non-extensive entropy formulations, such as Barrow, Rényi, and Sharma-Mittal entropy, within two frameworks: bulk boundary and RPS thermodynamics.
We first explore the thermodynamic topology in the bulk boundary framework. The illustrations are divided, with normalized field lines shown on the right. Figs.(\ref{m1}), (\ref{m2}), and (\ref{m3}) display the results for Barrow, Rényi, and Sharma-Mittal entropy, respectively. Figs. (\ref{1b}), (\ref{1d}), and (\ref{1h}) reveal three zero points, indicating topological charges determined by the free parameters and the non-extensive parameter $\delta$. These charges, which correlate with the winding number, are located within the blue contour loops at coordinates $(r, \theta)$. The sequence of these illustrations is governed by the parameter $\delta$.

The findings from these figures highlight a distinctive feature: three topological charges $(\omega = +1, -1, +1)$ and the total topological charge $W = +1$, represented by the zero points enclosed within the contour. Our analysis examines black hole stability by evaluating the winding numbers. Positive winding numbers suggest the thermodynamic stability of the on-shell black hole.\\

Additionally, as shown in Fig. (\ref{1f}), when the parameter $\delta$ increases to 0.9, the classification changes, and we observe two topological charges $(\omega = +1, -1)$ with a total topological charge $W = 0$. Also, as shown in Fig. (\ref{1h}), when the parameter $\delta$ is set to zero, our equations reduce to the Bekenstein-Hawking entropy structure, yielding the same results as in Figs. (\ref{1b}) and (\ref{1d}). Fig. (\ref{1h}) shows three topological charges $(\omega = +1, -1, +1)$ with a total topological charge $W = +1$.

\subsubsection{Thermodynamic topology within Rényi statistics}
\begin{figure}[h!]
 \begin{center}
 \subfigure[]{
 \includegraphics[height=4cm,width=4cm]{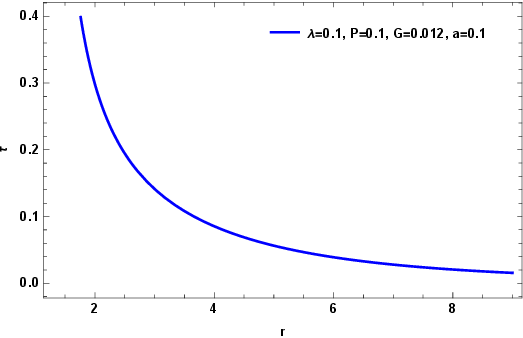}
 \label{2a}}
 \subfigure[]{
 \includegraphics[height=4cm,width=4cm]{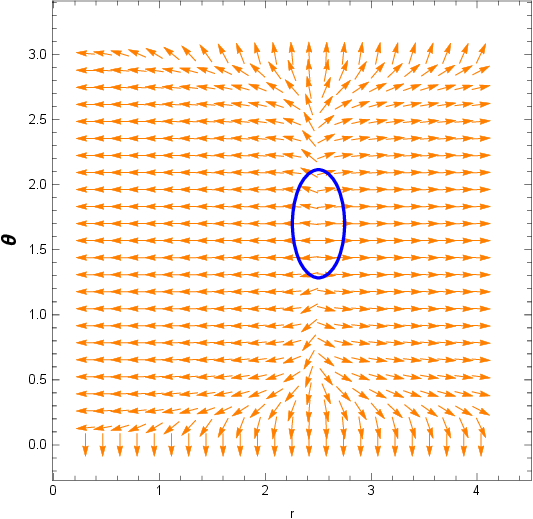}
 \label{2b}}
 \subfigure[]{
 \includegraphics[height=4cm,width=4cm]{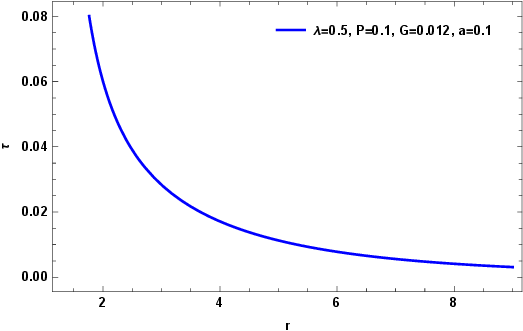}
 \label{2c}}
 \subfigure[]{
 \includegraphics[height=4cm,width=4cm]{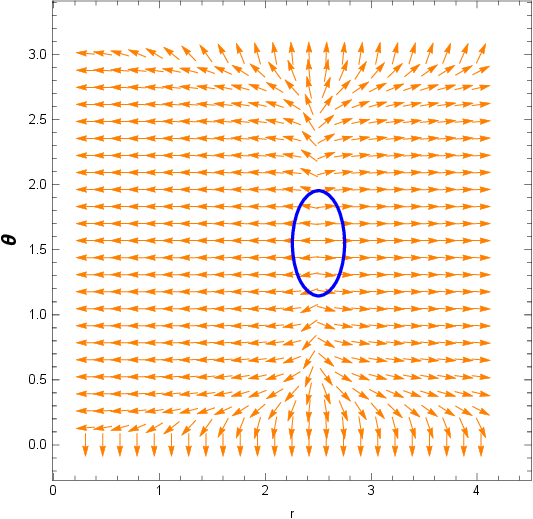}
 \label{2d}}\\
 \subfigure[]{
 \includegraphics[height=4cm,width=4cm]{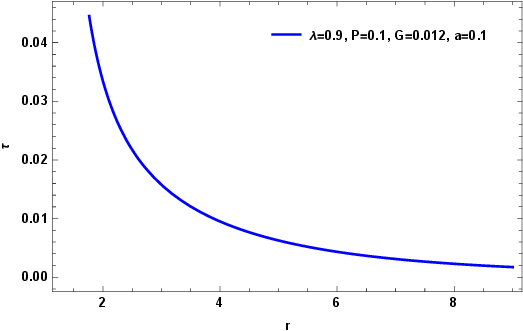}
 \label{2e}}
 \subfigure[]{
 \includegraphics[height=4cm,width=4cm]{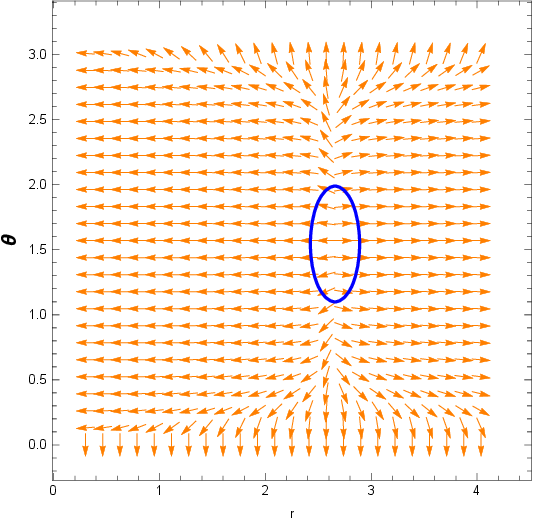}
 \label{2f}}
 \subfigure[]{
 \includegraphics[height=4cm,width=4cm]{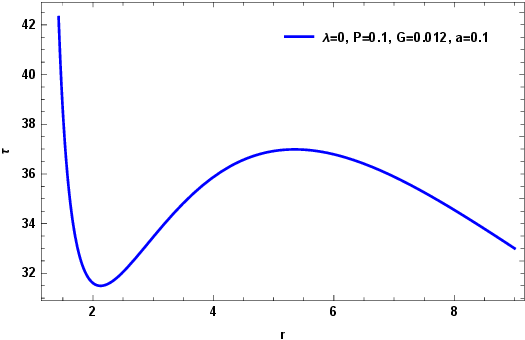}
 \label{2g}}
 \subfigure[]{
 \includegraphics[height=4cm,width=4cm]{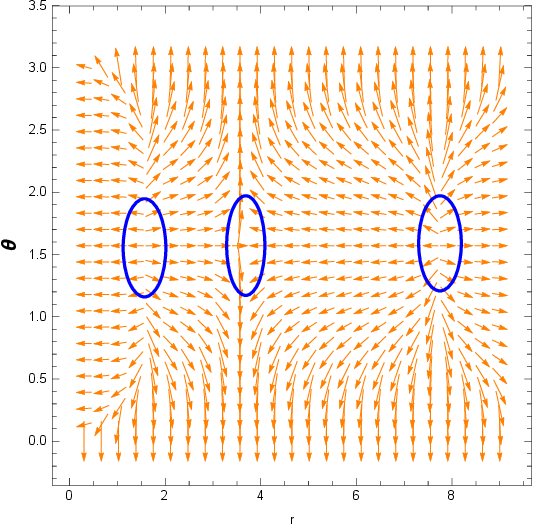}
 \label{2h}}
  \caption{\small{The curve described by Eq. (\ref{MBR2}) is illustrated in Figs. (\ref{2a}), (\ref{2c}), (\ref{2e}), and (\ref{2g}). In Figs. (\ref{2b}), (\ref{2d}), (\ref{2f}), and (\ref{2h}), the zero points (ZPs) are located at coordinates $(r, \theta)$ on the circular loops, corresponding to the nonextensive parameter $\lambda$.}}
 \label{m2}
 \end{center}
 \end{figure}

We also extend our study to include Rényi entropy within the context of bulk boundary thermodynamics. By utilizing Eqs. (\ref{N2}), (\ref{F1}), and (\ref{MB1}), we derive the function $\mathcal{F}$. Consequently, we can calculate $\phi^{r_h}$ with respect to Eqs. (\ref{F2}) as follows,
\begin{equation}\label{MBR1}
\begin{split}
\phi^r=-\frac{\frac{a-8 \pi  G P r^4-r^2}{G}+\frac{4 \pi  r \left(2 a G+r^2\right)}{\tau  \left(4 \pi  a G \lambda  \ln \left(\frac{r}{\sqrt{a}}\right)+G+\pi  \lambda  r^2\right)}+Q^2}{2 r^2}
\end{split}
\end{equation}
Furthermore, we determine $\tau$ as follows,
\begin{equation}\label{MBR2}
\tau =-\frac{4 \pi  \left(2 a G^2 r+G r^3\right)}{\left(a-8 \pi  G P r^4+G Q^2-r^2\right) \left(4 \pi  a G \lambda  \ln \left(\frac{r}{\sqrt{a}}\right)+G+\pi  \lambda  r^2\right)}
\end{equation}
Fig. (\ref{m2}) shows the results for Rényi entropy. As seen in Fig. (\ref{m2}), by setting the parameter $\lambda$ to zero, the number of topological charges increases $(\omega = +1, -1, +1)$ with the total topological charge $W = +1$. The number of total topological charges in the Bekenstein-Hawking entropy differs with the presence of the Rényi non-extensive parameter. We encounter a single topological charge $(\omega = +1)$ with the total topological charge $W = +1$, as shown in Figs. (\ref{2b}), (\ref{2d}), and (\ref{2f}).

\subsubsection{Thermodynamic topology within Sharma-Mittal statistics}
We further extend our study to incorporate Sharma-Mittal entropy within the framework of bulk boundary thermodynamics. By employing Eqs. (\ref{N3}), (\ref{F1}), and (\ref{MB1}), we derive the function $\mathcal{F}$. Also, we can calculate $\phi^{r_h}$ using Eqs. (\ref{F2}) as follows,
\begin{equation}\label{MBSM1}
\phi^r=-\frac{\frac{a-8 \pi  G P r^4-r^2}{G}+\frac{4 \pi  r \left(2 a G+r^2\right) \left(4 \pi  a \beta  \ln \left(\frac{r}{\sqrt{a}}\right)+\frac{\pi  \beta  r^2}{G}+1\right)^{\alpha /\beta }}{\tau  \left(4 \pi  a \beta  G \ln \left(\frac{r}{\sqrt{a}}\right)+G+\pi  \beta  r^2\right)}+Q^2}{2 r^2}
\end{equation}
Here, we calculate $\tau$ as follows,
\begin{equation}\label{MBSM2}
\tau =\frac{4 \pi  G r \left(2 a G+r^2\right) \left(4 \pi  a \beta  \log \left(\frac{r}{\sqrt{a}}\right)+\frac{\pi  \beta  r^2}{G}+1\right)^{\alpha /\beta }}{\left(-a+8 \pi  G P r^4-G Q^2+r^2\right) \left(4 \pi  a \beta  G \ln \left(\frac{r}{\sqrt{a}}\right)+G+\pi  \beta  r^2\right)}
\end{equation}
Fig. (\ref{m3}) illustrates Sharma-Mittal entropy with different classifications and the number of topological charges influenced by the Sharma-Mittal entropy non-extensive parameters $\alpha$ and $\beta$. As shown in Fig. (\ref{m3}), when the non-extensive parameters $\alpha$ and $\beta$ have values close to each other, they exhibit three topological charges $(\omega = +1, -1, +1)$ with a total topological charge $W = +1$. However, by keeping one parameter constant and varying the other, both the number of topological charges and the topological classification change completely. These changes in topological charges and classification are evident in Fig. (\ref{m3}). This demonstrates that the classification and number of topological charges are significantly influenced by the non-extensive parameters $\alpha$ and $\beta$, highlighting the importance of using these non-extensive entropies compared to the usual Bekenstein-Hawking case.
\begin{figure}[h!]
 \begin{center}
 \subfigure[]{
 \includegraphics[height=4cm,width=4cm]{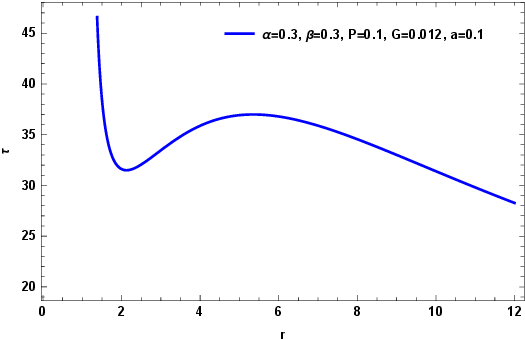}
 \label{3a}}
 \subfigure[]{
 \includegraphics[height=4cm,width=4cm]{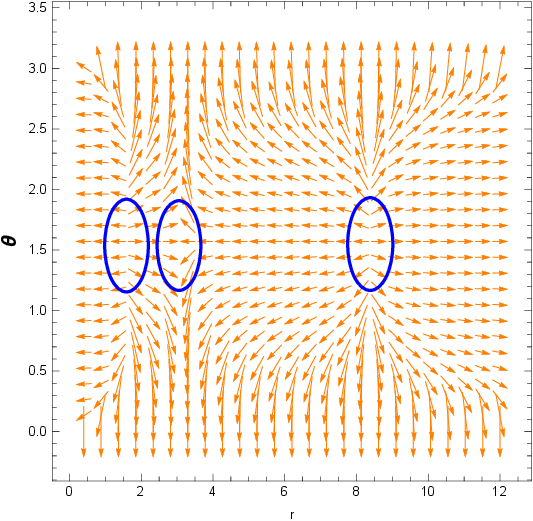}
 \label{3b}}
 \subfigure[]{
 \includegraphics[height=4cm,width=4cm]{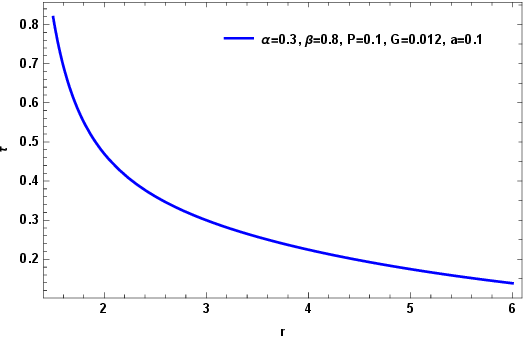}
 \label{3c}}
 \subfigure[]{
 \includegraphics[height=4cm,width=4cm]{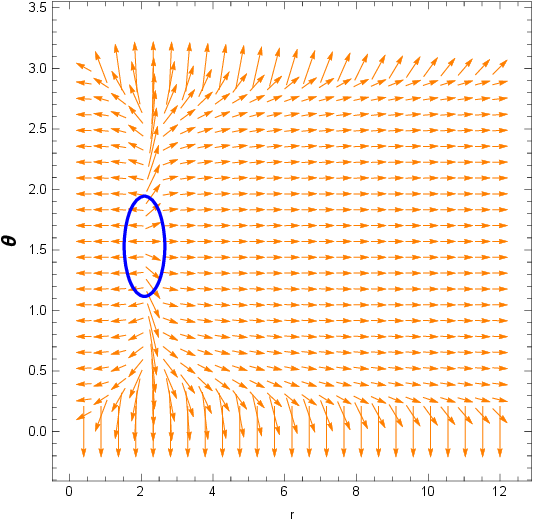}
 \label{3d}}\\
 \subfigure[]{
 \includegraphics[height=4cm,width=4cm]{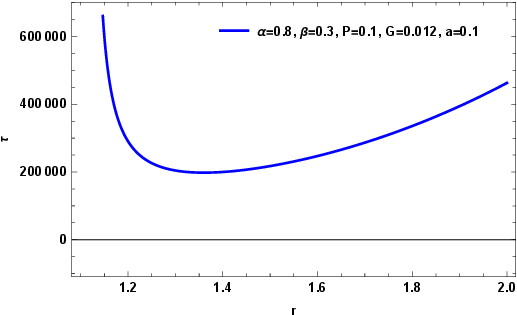}
 \label{3e}}
 \subfigure[]{
 \includegraphics[height=4cm,width=4cm]{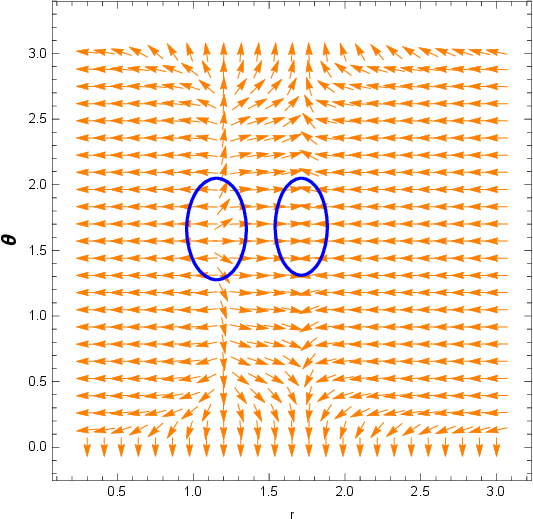}
 \label{3f}}
 \subfigure[]{
 \includegraphics[height=4cm,width=4cm]{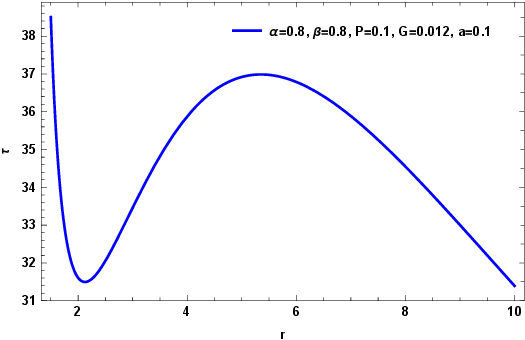}
 \label{3g}}
 \subfigure[]{
 \includegraphics[height=4cm,width=4cm]{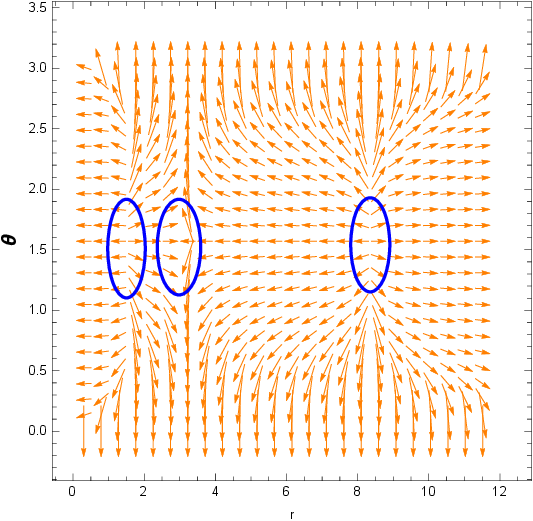}
 \label{3h}}\\
 \subfigure[]{
 \includegraphics[height=4cm,width=4cm]{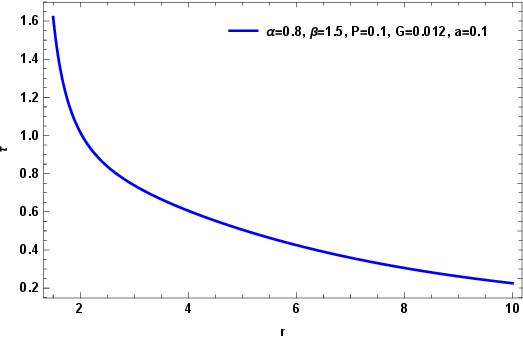}
 \label{3i}}
 \subfigure[]{
 \includegraphics[height=4cm,width=4cm]{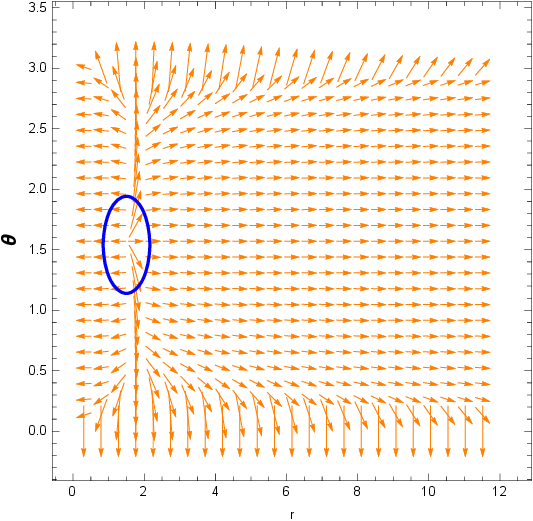}
 \label{3j}}
 \subfigure[]{
 \includegraphics[height=4cm,width=4cm]{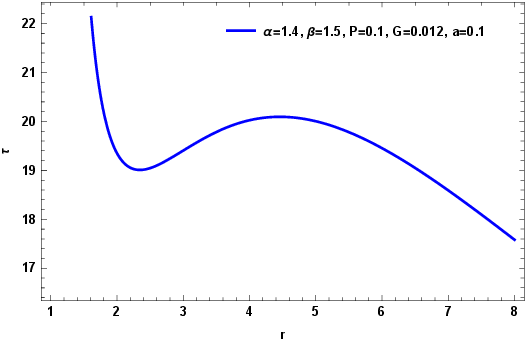}
 \label{3k}}
 \subfigure[]{
 \includegraphics[height=4cm,width=4cm]{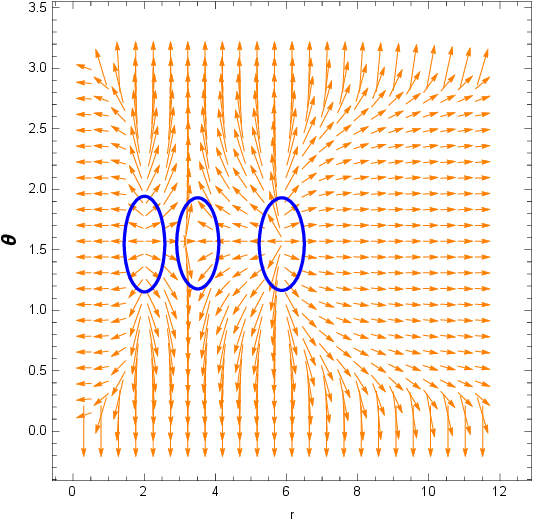}
 \label{3l}}
  \caption{\small{The curve described by Eq. (\ref{MBSM2}) is illustrated in Figs. (\ref{3a}), (\ref{3c}), (\ref{3e}), (\ref{3g}), and (\ref{3i}). In Figs. (\ref{3b}), (\ref{3d}), (\ref{3f}), (\ref{3h}), and (\ref{3j}), the zero points (ZPs) are located at coordinates $(r, \theta)$ on the circular loops, corresponding to the nonextensive parameters $(\alpha)$ and $(\beta)$.}}
 \label{m3}
 \end{center}
 \end{figure}
 \newpage
\subsection{RPS thermodynamics}
In this section, we will rederive the equations for a 4D AdS Einstein-Gauss-Bonnet (EGB) black hole. The entropy, based on above equation in RPS thermodynamics, are given by,
\begin{equation}\label{MRPS1}
q = \frac{\hat{q}}{\sqrt{C}}, \quad S = \frac{C r_h^2 \pi}{l^2} + 4 \ln \left(\frac{r_h }{\sqrt{a}}\right)a \pi, \quad G = \frac{l^2}{C}
\end{equation}
The temperature $T$ is expressed as,
\begin{equation}\label{MRPS2}
T = \frac{-\frac{\hat{q}^2 l^4}{C^2} + l^2 r_h^2 + 3r_h^4 - l^2 a}{4 (r_h^2 + 2a) r_h l^2 \pi}.
\end{equation}
The parameter $C$ is determined by,
\begin{equation}\label{MRPS3}
C = \frac{\sqrt{-(-l^2 r_h^4 + 3r_h^6 + 5l^2 a r_h^2 + 18 a r_h^4 + 2 a^2 l^2)(3r_h^2 + 2a)}l^2\hat{q}}{- l^2 r_h^4 + 3r_h^6 + 5l^2 a r_h^2 + 18 a r_h^4 + 2 a^2 l^2}
\end{equation}
\subsubsection{Thermodynamic topology within Barrow statistics}
Similar to the previous section, we can study the incorporation of non-extensive entropy within the framework of RPS thermodynamics. For Barrow entropy, by employing Eqs. (\ref{N1}), (\ref{F1}), and (\ref{MRPS1}), we derive the function $\mathcal{F}$. Subsequently, we calculate $\phi^{r_h}$ using Eqs. (\ref{F2}) as follows,
\begin{equation}\label{MRPSB1}
\phi^r=\frac{\tau  \left(-a C^2 l^2+C^2 l^2 r^2+3 C^2 r^4+l^4 \left(-\hat{q}^2\right)\right)-2 \pi  C (\delta +2) l^2 r \left(2 a l^2+C r^2\right) \left(4 \pi  a \ln \left(\frac{r}{\sqrt{a}}\right)+\frac{\pi  C r^2}{l^2}\right)^{\delta /2}}{2 C l^4 r^2 \tau }
\end{equation}
The $(\tau)$ is obtained as follows,
\begin{equation}\label{MRPSB2}
\tau =\frac{2 \pi  C (\delta +2) l^2 r \left(2 a l^2+C r^2\right) \left(4 \pi  a \ln \left(\frac{r}{\sqrt{a}}\right)+\frac{\pi  C r^2}{l^2}\right)^{\delta /2}}{-a C^2 l^2+C^2 l^2 r^2+3 C^2 r^4+l^4 \left(-\hat{q}^2\right)}
\end{equation}
A particularly intriguing aspect of this study is its extension to the restricted phase space (RPS). When we continue our investigations in this space using the two mentioned entropies, we observe that, under all conditions and for all free parameters, the topological charge consistently remains $(\omega = +1)$ with a total topological charge $W = +1$. This consistency indicates a stable topological structure within the RPS framework, regardless of the specific values of the free parameters.

\begin{figure}[h!]
 \begin{center}
 \subfigure[]{
 \includegraphics[height=3cm,width=2.8cm]{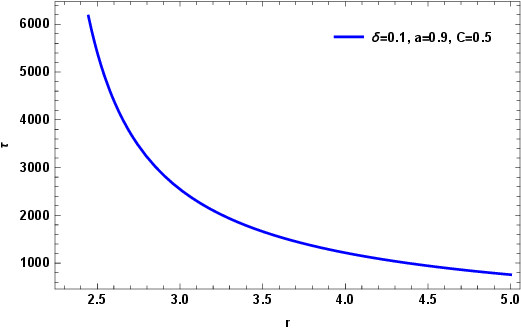}
 \label{4a}}
 \subfigure[]{
 \includegraphics[height=3cm,width=2.8cm]{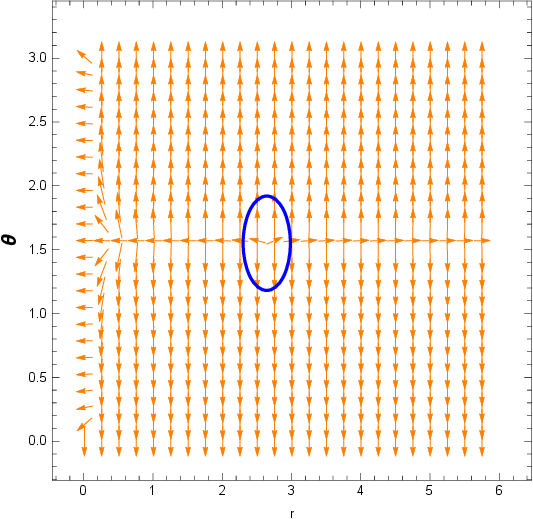}
 \label{4b}}
 \subfigure[]{
 \includegraphics[height=3cm,width=2.8cm]{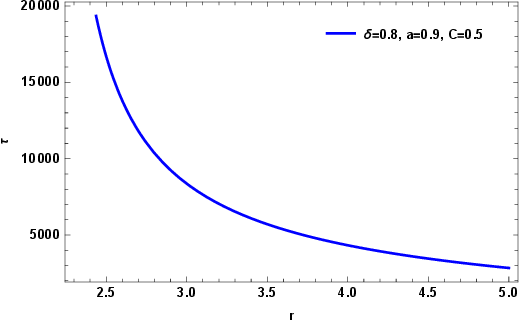}
 \label{4c}}
 \subfigure[]{
 \includegraphics[height=3cm,width=2.8cm]{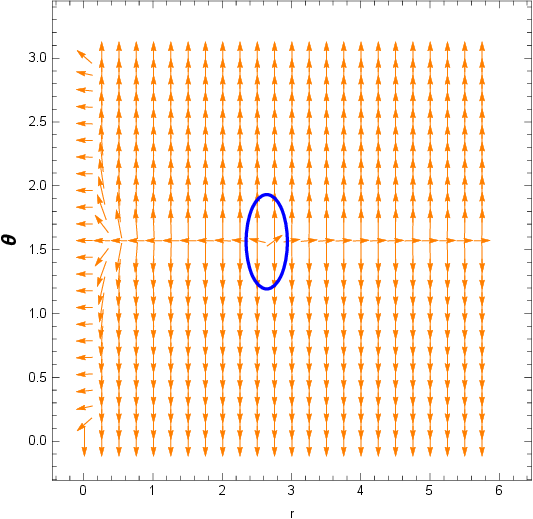}
 \label{4d}}
 \subfigure[]{
 \includegraphics[height=3cm,width=2.8cm]{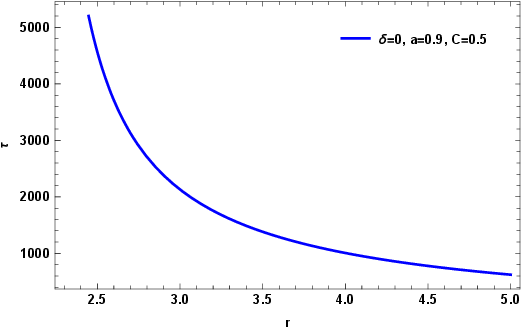}
 \label{4e}}
 \subfigure[]{
 \includegraphics[height=3cm,width=2.8cm]{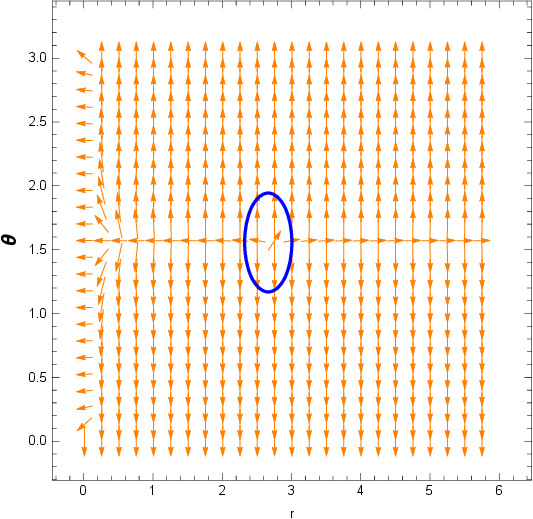}
 \label{4f}}
  \caption{\small{The curve described by Eq. (\ref{MRPSB2}) is illustrated in Figs. (\ref{4a}), (\ref{4c}), and (\ref{4e}). In Figs. (\ref{4b}), (\ref{4d}), and (\ref{4f}), the zero points (ZPs) are located at coordinates $(r, \theta)$ on the circular loops, corresponding to the nonextensive parameter $\delta$.}}
 \label{m4}
 \end{center}
 \end{figure}

\begin{figure}[h!]
 \begin{center}
 \subfigure[]{
 \includegraphics[height=3cm,width=2.8cm]{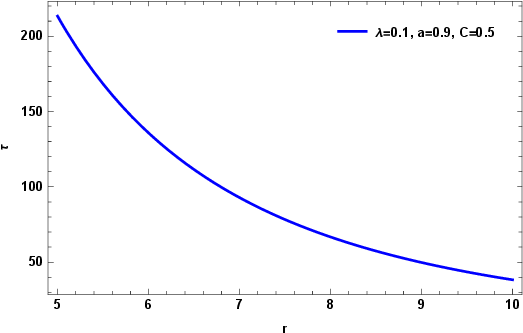}
 \label{5a}}
 \subfigure[]{
 \includegraphics[height=3cm,width=2.8cm]{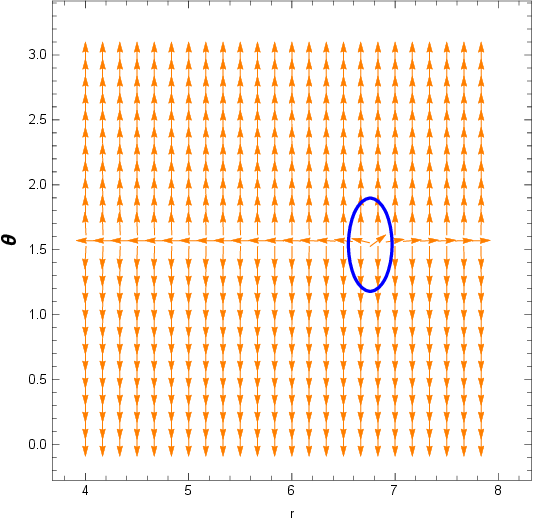}
 \label{5b}}
 \subfigure[]{
 \includegraphics[height=3cm,width=2.8cm]{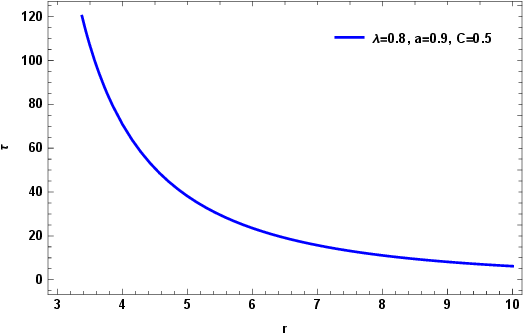}
 \label{5c}}
 \subfigure[]{
 \includegraphics[height=3cm,width=2.8cm]{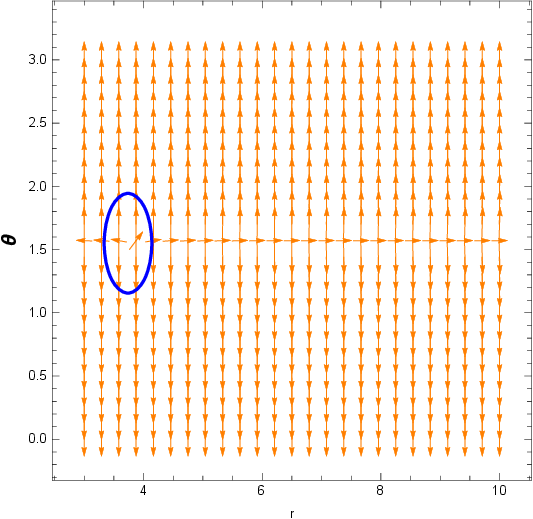}
 \label{5d}}
 \subfigure[]{
 \includegraphics[height=3cm,width=2.8cm]{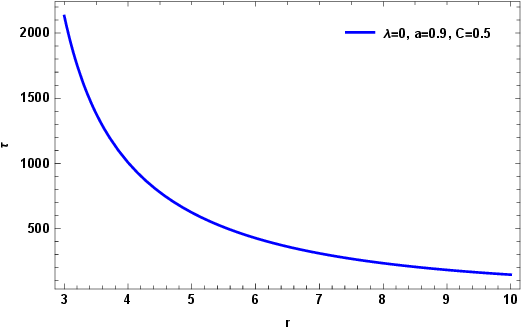}
 \label{5e}}
 \subfigure[]{
 \includegraphics[height=3cm,width=2.8cm]{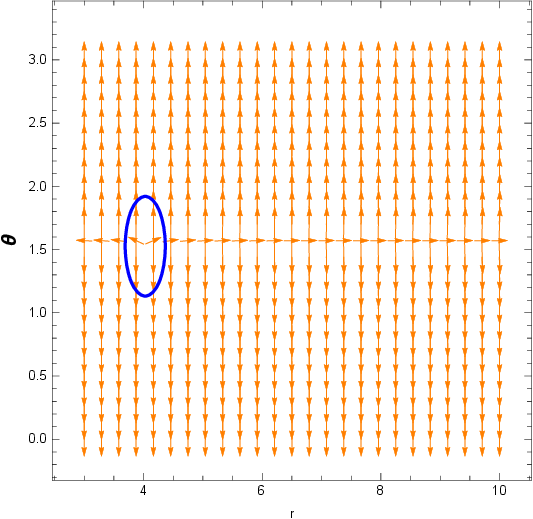}
 \label{5f}}
  \caption{\small{The curve described by Eq. (\ref{MRPSR2}) is illustrated in Figs. (\ref{5a}), (\ref{5c}), and (\ref{5e}). In Figs. (\ref{5b}), (\ref{5d}), and (\ref{5f}), the zero points (ZPs) are located at coordinates $(r, \theta)$ on the circular loops, corresponding to the nonextensive parameter $(\lambda)$.}}
 \label{m5}
 \end{center}
 \end{figure}

Additionally, when we reduce the analysis to Bekenstein-Hawking entropy within RPS, we observe similar behavior. This suggests that, unlike in the bulk boundary space, the RPS framework exhibits a uniform topological behavior across both non-extensive entropy and Hawking entropy states. This uniformity is illustrated in Figs. (\ref{m4}), (\ref{m5}), and (\ref{m6}), where the topological charges and their configurations remain consistent.

\subsubsection{Thermodynamic topology within Rényi statistics}
For Rényi entropy, by utilizing Eqs. (\ref{N2}), (\ref{F1}), and (\ref{MRPS1}), we derive the function $\mathcal{F}$. Subsequently, we calculate $\phi^{r_h}$ using Eqs. (\ref{F2}) as follows,
\begin{equation}\label{MRPSR1}
\phi^r=-\frac{\frac{4 \pi  r \left(2 a l^2+C r^2\right)}{\tau  \left(4 \pi  a \lambda  l^2 \ln \left(\frac{r}{\sqrt{a}}\right)+\pi  C \lambda  r^2+l^2\right)}+\frac{a C}{l^2}-\frac{C r^2 \left(l^2+3 r^2\right)}{l^4}+\frac{\hat{q}^2}{C}}{2 r^2}
\end{equation}
Also, we have,
\begin{equation}\label{MRPSR2}
\tau =-\frac{4 \pi  \left(2 a C l^6 r+C^2 l^4 r^3\right)}{\left(a C^2 l^2-C^2 l^2 r^2-3 C^2 r^4+l^4 \hat{q}^2\right) \left(4 \pi  a \lambda  l^2 \ln \left(\frac{r}{\sqrt{a}}\right)+\pi  C \lambda  r^2+l^2\right)}
\end{equation}
The implications of these findings are significant. They suggest that the RPS framework provides a robust and stable environment for studying the thermodynamic properties of black holes, irrespective of the entropy model used. This stability is crucial for understanding the fundamental nature of black hole thermodynamics and could provide insights into the behavior of black holes under various theoretical models.

\subsubsection{Thermodynamic topology within Sharma-Mittal statistics}
Here, with respect to Eqs. (\ref{N3}), (\ref{F1}), and (\ref{F2}), the $\phi^{r_h}$ is calculated for Sharma-Mittal entropy in RPS thermodynamics as follows,
\begin{equation}\label{MRPSSM1}
\phi^r=-\frac{\frac{C^2 \left(l^2 \left(a+r^2\right)+r^4\right)+l^4 \hat{q}^2}{C r^2}+\frac{4 \pi  l^2 \left(2 a l^2+C r^2\right) \left(4 \pi  a \beta  \ln \left(\frac{r}{\sqrt{a}}\right)+\frac{\pi  \beta  C r^2}{l^2}+1\right)^{\frac{\alpha }{\beta }-1}}{r \tau }-2 C \left(l^2+2 r^2\right)}{2 l^4}
\end{equation}
Then we can calculate the $\tau$,
\begin{equation}\label{MRPSSM2}
\tau =\frac{4 \pi  C l^4 r \left(2 a l^2+C r^2\right) \left(4 \pi  a \beta  \ln \left(\frac{r}{\sqrt{a}}\right)+\frac{\pi  \beta  C r^2}{l^2}+1\right)^{\alpha /\beta }}{\left(-a C^2 l^2+C^2 l^2 r^2+3 C^2 r^4+l^4 \left(-\hat{q}^2\right)\right) \left(4 \pi  a \beta  l^2 \ln \left(\frac{r}{\sqrt{a}}\right)+\pi  \beta  C r^2+l^2\right)}
\end{equation}

\begin{figure}[h!]
 \begin{center}
 \subfigure[]{
 \includegraphics[height=3cm,width=2.8cm]{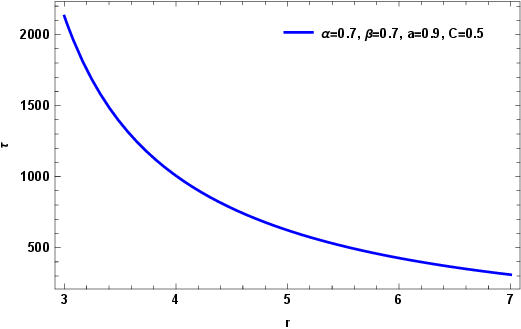}
 \label{6a}}
 \subfigure[]{
 \includegraphics[height=3cm,width=2.8cm]{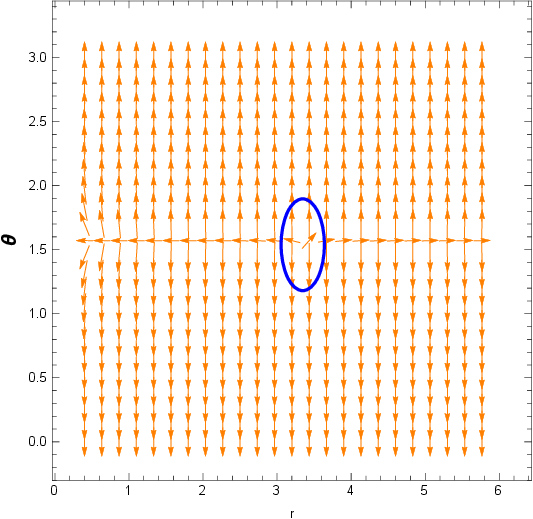}
 \label{6b}}
 \subfigure[]{
 \includegraphics[height=3cm,width=2.8cm]{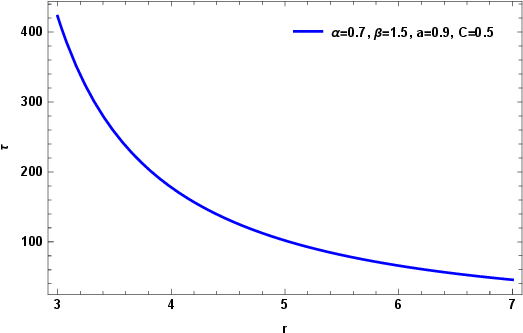}
 \label{6c}}
 \subfigure[]{
 \includegraphics[height=3cm,width=2.8cm]{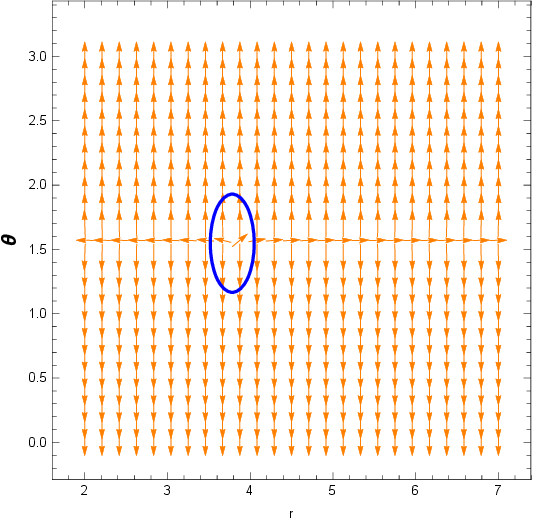}
 \label{6d}}
 \subfigure[]{
 \includegraphics[height=3cm,width=2.8cm]{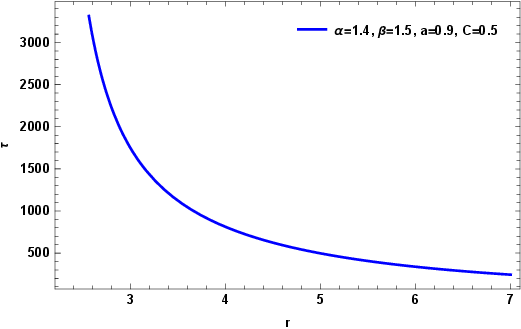}
 \label{6e}}
 \subfigure[]{
 \includegraphics[height=3cm,width=2.8cm]{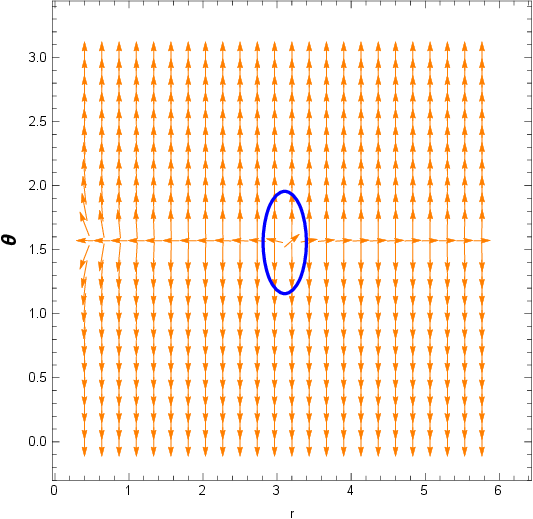}
 \label{6f}}
  \caption{\small{The curve described by Eq. (\ref{MRPSSM2}) is illustrated in Figs. (\ref{6a}), (\ref{6c}), and (\ref{6e}). In Figs. (\ref{6b}), (\ref{6d}), and (\ref{6f}), the zero points (ZPs) are located at coordinates $(r, \theta)$ on the circular loops, corresponding to the nonextensive parameters $(\alpha)$ and $(\beta)$}}
 \label{m3}
 \end{center}
 \end{figure}
Moreover, the consistency in topological charges within RPS, as opposed to the variability observed in the bulk boundary space, highlights the potential advantages of using RPS for such studies. It underscores the importance of considering different frameworks and entropy models to gain a comprehensive understanding of black hole thermodynamics.

In summary, the extension of our study to the restricted phase space reveals a stable and consistent topological structure across different entropy models, providing valuable insights into the thermodynamic behavior of black holes. This consistency, illustrated in the figures, emphasizes the robustness of the RPS framework in capturing the essential features of black hole thermodynamics.
\section{Conclusion}
In this paper, we explore the thermodynamic topology of AdS Einstein-Gauss-Bonnet black holes using non-extensive entropy formulations, including Barrow, Rényi, and Sharma-Mittal entropy, within two distinct frameworks: bulk boundary and restricted phase space (RPS) thermodynamics.

Our findings in the bulk boundary framework reveal significant variability in topological charges influenced by the free parameters and the Barrow non-extensive parameter $(\delta)$. Specifically, we identify three topological charges $(\omega = +1, -1, +1)$. When the parameter $(\delta)$ increases to 0.9, the classification changes, resulting in two topological charges $(\omega = +1, -1)$. When $(\delta)$ is set to zero, the equations reduce to the Bekenstein-Hawking entropy structure, yielding consistent results with three topological charges.

Additionally, setting the non-extensive parameter $(\lambda)$ in Rényi entropy to zero increases the number of topological charges, but the total topological charge remains $(W = +1)$. The presence of the Rényi non-extensive parameter alters the topological behavior compared to the Bekenstein-Hawking entropy. Sharma-Mittal entropy shows different classifications and various numbers of topological charges influenced by the non-extensive parameters $(\alpha)$ and $(\beta)$. When $(\alpha)$ and $(\beta)$ have values close to each other, three topological charges with a total topological charge $(W = +1)$ are observed. Varying one parameter while keeping the other constant significantly changes the topological classification and number of topological charges.

In contrast, the RPS framework demonstrates remarkable consistency in topological behavior. Under all conditions and for all free parameters, the topological charge remains $(\omega = +1)$ with a total topological charge $(W = +1)$. This uniformity persists even when reduced to Bekenstein-Hawking entropy, suggesting that the RPS framework provides a stable environment for studying black hole thermodynamics across different entropy models.

These findings underscore the importance of considering various entropy formulations and frameworks to gain a comprehensive understanding of black hole thermodynamics. The variability observed in the bulk boundary framework highlights the dynamic nature of topological charges influenced by different parameters, while the consistency in the RPS framework emphasizes its robustness and stability. This dual approach provides valuable insights into the fundamental nature of black hole thermodynamics and the stability of black holes under different theoretical models. By exploring these diverse entropy formulations, we can better understand the intricate behaviors and properties of black holes, paving the way for future research in this fascinating field.
We face some questions that highlight potential avenues for future research, aiming to deepen our understanding of black hole thermodynamics and the role of nonextensive entropy in this fascinating area of study.\\
1.  How do different values of the nonextensive parameters $(\lambda)$, $(\alpha)$, and $(\beta)$ affect the stability and phase transitions of black holes in other types of spacetimes?\\
2.  How is behavior the analysis of the thermodynamic topology in higher-dimensional spacetimes with nonextensive entropy?\\
3. What influences our understanding of black hole entropy in the context of quantum gravity theories?\\
4. Is there a critical value of the nonextensive parameters beyond which the thermodynamic behavior of black holes significantly deviates from the predictions of classical thermodynamics?\\
5. How can the consistency of topological charges in restricted phase space be leveraged to develop new models for black hole thermodynamics?
6. What experimental or observational evidence could be used to validate the theoretical predictions made using nonextensive entropy frameworks in black hole thermodynamics?

\end{document}